\documentstyle[preprint,aps,psfig,pra]{revtex}
\draft
\tightenlines
\begin{document}
\title{Development of an approximation scheme for quasi-exactly solvable
double-well potentials}
 \author{R. Atre \thanks{atre@prl.res.in} $^{1,2}$
 and P. K. Panigrahi \thanks{prasanta@prl.res.in} $^{1,2}$}
 \address{$^1$ Physical Research Laboratory,
 Navrangpura, Ahmedabad-380 009, India\\
 $^2$ School of Physics, University of Hyderabad, Hyderabad-500 046, India}
 %\date{\today}
 \maketitle

%**************************** ABSTRACT************************************
\begin{abstract}
We make use of a recently developed method to, not only obtain the exactly
known eigenstates and eigenvalues of a number of
quasi-exactly solvable Hamiltonians, but also construct a convergent
approximation scheme for locating those levels, not amenable to analytical
treatments. The fact that, the above method yields an expansion of the
wave functions in terms of corresponding energies, enables one to treat energy
as a variational parameter, which can be effectively used for the identification of
the eigenstates. It is particularly useful for the quasi-exactly solvable
systems, where the ground state is known and  a number of
eigenstates are bounded, both below and above. The efficacy of the procedure
is illustrated by obtaining, the low-lying excited states of a prototypical
double-well potential, where the conventional techniques are not very reliable.
Our approach yields the approximate eigenfunctions and eigenvalues, whose
accuracy can be improved to any desired level, in a controlled manner.
Comparing the present results with those of an independent numerical
method, it was found that, the first few terms in our approximate solutions are
enough to yield the excited state eigenvalues, accurate upto the third place
of the decimal.

\end{abstract}
%*********************************** INTRODUCTION ******************************
\newpage
\section{Introduction}
Quasi-exactly solvable (QES) systems are intermediate to exactly solvable and
non-solvable ones. These quantum systems are characterized by the fact that,
only a finite number of the eigenstates can be analytically determined
\cite{Singh,Turb}. One dimensional examples include anharmonic potentials,
some of which have a double-well structure. Interestingly, a QES potential with
a centrifugal barrier appears in the treatment of ring lasers, where
the Fokker-Planck equation is converted to an equivalent Schr$\ddot{o}$dinger
eigenvalue problem \cite{Wolf}. Other QES examples have been encountered in similar
problems \cite{Kampen,Brown}. These type of Hamiltonians,
amenable to partial algebraization, have also manifested in various
other areas of physics \cite{Shifm} and have attracted considerable attention
in the recent literature \cite{Ushve,Khare}. A
number of independent methods, from group theoretical to
algebraic ones, have been developed for studying the QES systems \cite{Turb}.

In the conventional approach to quasi-exactly solvable systems,
one starts with an operator suitably constructed from the
differential realizations of the generators of a given Lie
algebra, acting in the finite dimensional space of monomials. A
QES Hamiltonian is then arrived at by converting the above problem
into a Schr$\ddot{o}$dinger eigenvalue equation, through
appropriate similarity transformations. Although much effort has
gone into studying the structure of the analytically available
part of the spectrum, there has not been any attempt in the
literature, to the best of the authors' knowledge, to find the
ones, not determined analytically. The fact that, some of the QES
potentials exhibit a double-well structure, for which determining
the approximate eigenvalues and eigenstates have been quite
challenging, makes this problem worth investigating. Further, a number of
these states are constrained to lie between certain range of
energies; hence, apart from academic interest, developing an
approximation scheme to locate these eigenvalues and
eigenfunctions, provides an ideal ground for testing the proposed
scheme. As mentioned earlier, a number of these type
of potentials appear in the Schr$\ddot{o}$dinger equation
originating from the Fokker-Planck equation, governing the
dynamics of non-equilibrium systems. Finding the low-lying states
of these problems accurately have physical significance; for example the 
tunneling rate is related to the energy difference between the ground and 
first excited state \cite{Brown,Kumar,Sukhatme}.

The goal of this paper is to develop an accurate approximation scheme for
finding the non-exactly determined eigenstates of the QES Hamiltonians.
For this purpose, we employ a
recently developed method for solving linear differential
equations \cite{pani2,pkp,charan}, used earlier for diagonalizing many-body interacting
systems \cite{pkp,prb}. The analytically obtainable part of the spectra of a
number of QES systems is first determined, in the following section, to
demonstrate the working of the
above method. We then proceed, in Sec.III, to compute approximately,
those eigenvalues and eigenstates of a prototype QES system, which are not 
amenable to analytical treatment. The fact that our procedure yields a series 
solution for a differential equation in terms of energy, allows us to treat 
the same as a variational parameter. This can be effectively used for the 
identification of the eigenstates and corresponding eigenvalues, as will be 
shown in the text. It is particularly useful for the quasi-exactly solvable 
systems, where the ground state is known and a number of eigenstates are both 
bounded below and above.

The example of a double-well potential is taken deliberately, since the
conventional techniques are not very reliable for the same. The usefulness of
the present approach is shown by computing a number of low-lying eigenstates.
Our approach yields the approximate eigenfunctions and
eigenvalues, whose accuracy can be improved to any desired level, in a
controlled manner. We then compare our results, with another convergent
numerical scheme \cite{Korsch,Sheorey}; it is found that, the first few terms in our approximate
solutions are enough to yield the
excited state eigenvalues, accurate upto the third place of the decimal. We conclude in Sec.IV, after 
pointing out the advantages and limitations of the present approximation scheme and directions for 
further investigations.

%************************ THE METHOD ******************************

\section{Exact eigenstates of quasi-exactly solvabe systems: a novel approach}
In this section, we obtain the analytically solvable part of the eigenspectra,
of a class of
QES systems, making use of a recently developed method for solving linear
differential equations \cite{pani2}. In this procedure, the solution space of the
differential equation is connected with the space of monomials.
We concentrate on the QES systems having polynomial potentials, with or
without a centrifugal barrier term, although the method can be applied to other
systems as well. As will become clear in the subsequent section, the same
procedure yields the approximate eigenfunctions and eigenvalues, for the
analytically inaccessible states of the QES systems, to the desired accuracy.

A single variable differential equation, after suitable
manipulations (which will become clear from the examples in the
text), can be written as,
\begin{equation} \label{ie}
\left[F(D) + P(x,d/dx)\right] y(x) = 0 \quad,
\end{equation}
where, $D \equiv x \frac{d}{dx}$ is the Euler operator, $F(D)
\equiv \sum_{n = - \infty}^{n = \infty} a_n D^n $ and $a_n$'s are
some parameters; $P(x,d/dx)$ can be an arbitrary polynomial
function of $x$, $\frac{d}{dx}$ and other operators. It can be
straightforwardly shown by direct substitution that,
the solution to Eq. (\ref{ie}) can be written in the form,
\begin{eqnarray} \label{an}
y(x) &=& C_\lambda \left \{\sum_{m = 0}^{\infty} (-1)^m
\left[\frac{1}{F(D)}P(x,d/dx)\right]^m \right \} x^\lambda
\end{eqnarray}
provided, $F(D) x^\lambda = 0$ and the coefficient of $x^\lambda$
in $y(x) - C_\lambda x^\lambda$ is zero (no summation over
$\lambda$); here, $C_\lambda$ is a constant.
    This straightforward method, not only yields solutions to the
familiar differential equations \cite{charan}, but also leads to
the diagonalization of a number of correlated many-body
Hamiltonians \cite{prb}.

 For the QES case, we first consider the example of
the sextic oscillator, whose
Hamiltonian (in the units $\hbar=2m=1$) is given by
\begin{equation}  H =
- \frac{d^2}{dx^2}+\alpha x^2 +\beta x^4 +\gamma x^6 \quad.
\end{equation} It is well-known that, this problem is QES, provided a
certain relation exists between the parameters $\alpha,\beta$ and
$\gamma$. Instead of postulating the same, we first illustrate how
this condition emerges naturally. Asymptotic analysis indicates a
measure of the form $\hat{\psi_0}\equiv e^{-(ax^2+bx^4)}$, with
the unknown parameters, $a$ and $b$ to be determined from the
Hamiltonian parameters $\alpha$, $\beta$ and $\gamma$. A
similarity transformation $\tilde{
H}={\hat{\psi_0}^{-1}}H\hat{\psi_0}$, yields;
\begin{eqnarray}
\tilde{ H}&=&-
\frac{d^2}{dx^2}+8bx^3\frac{d}{dx}+4ax\frac{d}{dx}+(\alpha-4a^2+12b)x^2
\nonumber \\
\label{dh} &&+(\beta-16ab)x^4+(\gamma-16b^2)x^6+2a \qquad .
\end{eqnarray}
Setting the coefficients of $x^4$ and $x^6$ equal to zero, one
obtains,
\begin{equation} a= \frac{\beta}{4{\sqrt \gamma}} \quad
{\mathrm and} \quad b= \frac{\sqrt{\gamma}}{4} \quad \quad .
\end{equation}
Assuming that the solution of the eigenvalue equation
$\tilde{H}P_n(x) = EP_n(x) $ is a polynomial, of degree $n$, one
observes that, the operators, $2{\sqrt \gamma} x^3\frac{d}{dx}$
and $(\alpha-\frac{\beta^2}{4\gamma}+3\sqrt{\gamma})x^2$ increase
the degree of $P_n(x)$ by two. Preserving the degree of the
polynomial leads to the above mentioned relationship between the
coupling parameters of the sextic oscillator \cite{Ushve} :
\begin{equation}\label{condi}\frac{1}{\sqrt\gamma}{\left(\frac{\beta^2}{4
\gamma}-\alpha\right)}=2n+3 \quad.\end{equation} It is worth
mentioning that, in exactly solvable problems, the differential
operators do not increase the degree of the polynomial and hence
no additional condition is required there.

 The eigenvalue
problem can be cast in a form as given in Eq. (\ref{ie})\,:
\begin{eqnarray} \label{sextic1}
{\left[D(D-1)-2{\sqrt \gamma} x^5\frac{d}{dx}-{\frac{\beta}{\sqrt
\gamma} x^3\frac{d}{dx}+{2n\sqrt\gamma
}x^4+\left(E-\frac{\beta}{2\sqrt{\gamma}}
\right)x^2}\right]}P_{n}(x)=0  \qquad.
\end{eqnarray}
The condition $F(D)x^{\lambda}=0$, yields ${\lambda}=0$ and $1$. It is easy to
see that the two values of $\lambda$ separate the Hilbert space into even and
odd sectors.
The solution, corresponding to the root $\lambda=0$, can be
expanded as,
\begin{eqnarray} \label{Even} P_{n}^{0}(x) &=& C_{0}
\sum^{\infty}_{m=0}(-1)^m \left[ { \frac{1}{D(D-1)}} \left (
\tilde
Ex^2-{\frac{\beta}{\sqrt{\gamma}}}{x^3}{\frac{d}{dx}}+2n{\sqrt\gamma}{x^4}
-2{\sqrt\gamma}{x^5}{\frac{d}{dx}} \right)\right]^m x^{0} \qquad
\nonumber \\
&=&\sum^{\infty}_{k=0}Q_{k}(\tilde{E}){\frac{x^{2k}}{2k!}} \qquad
,
\end{eqnarray}
 where $\tilde{E}=E-\beta/2\sqrt\gamma$ and  $Q_{k}(\tilde{E})$
 is an appropriate polynomial in energy.

Since the degree of the polynomial has been already fixed at $n$,
an even integer in this case, we impose the condition that the
series terminates at the desired point by putting the coefficient
of the subsequent term to zero. As will be explicitly seen, the
same coefficient appears as a factor in rest of the terms of the
series, thereby ensuring that the degree of the polynomial is
maintained. Writing $n=4j$, where $j$ can take semi-integer
values, one needs to put $Q_{2j+1}(\tilde{E})=0$ to obtain the
above result. This condition leads to $2j+1$ independent solutions
for energy and the corresponding eigenfunctions. We would like to
emphasize that, so far we have not assumed any additional property
of the spectral problem, for arriving at these results.

In similar fashion, one can easily incorporate the centrifugal barrier in to QES
problems. The Hamiltonian of the sextic oscillator for $\beta =0$, with a centrifugal barrier is given
by,
\begin{eqnarray}
\hat{H}=
-\frac{d^2}{dx^2}+\frac{\sigma}{x^2}+{\alpha}x^2+{\gamma}x^6 \quad ;
\end{eqnarray}
the corresponding measure is of the form  $ \psi_{0}
=x^{2l}\exp{(-{ax^4}/4)}$. Performing an appropriate similarity
transformation and setting the coefficients of $x^6$ and $1/x^2$
to zero, we obtain,

\begin{eqnarray} \label{centri}
\left[- \frac{d^2}{dx^2}+ {\frac{-4l}{x}}{\frac{d}{dx}}+
2{\sqrt\gamma}x^3\frac{d}{dx}
+\left(\alpha+3{\sqrt\gamma}+4l{\sqrt\gamma}\right )x^2
\right]P_{n}(x)= EP_{n}(x) \quad ,
\end{eqnarray}

with $a = \sqrt\gamma$ \quad and \quad $l =
{\frac{1}{4}}+{\frac{1}{2}}\sqrt{\frac{1}{4}+\sigma}$.

The condition that $P_n(x)$ is a polynomial of degree $n$,
leads to,
\begin{equation}
-\frac{\alpha}{4\sqrt\gamma}+\frac{1}{2}\sqrt{\frac{1}{4}+\sigma}=n/2+1
\quad .
\end{equation}
Multiplication of Eq. (\ref{centri}) with $-x^2$ yields,
\begin{eqnarray}
\left[D(D+4l-1)+Ex^2+2n\sqrt{\gamma}x^4
-2\sqrt{\gamma}x^5{\frac{d}{dx}}\right]P_{n}(x)=0 \qquad ,
\end{eqnarray}
which  $\lambda=0$ and $1-4l$, as the solutions of
$F(D)x^\lambda=0$. A  suitable parameterization in the form of
\begin{eqnarray}
 \alpha=-4a{\left(s+\frac{1}{2}+\mu\right)}
\mathrm{and}\quad
\sigma=4{\left(s-\frac{1}{4}\right)}{\left(s-\frac{3}{4}\right)}
\nonumber
\end{eqnarray} makes the result amenable for comparison with the
existing literature \cite{Ushve}. Analogous to the earlier example, one can now
find solutions corresponding to various values of $n$. For $n=2$
one finds,
\begin{eqnarray}
E_{2\pm}=\pm\sqrt{32as} \nonumber \\ {\mathrm{and}} \quad
P_{2\pm}(x)=\left[ax^2-\frac{E_{2\pm}}{4}\right] \quad .
\end{eqnarray}
Even states for other values of $n$ can also be similarly
obtained.

In the absence of a centrifugal barrier, for the anharmonic oscillator of
Eq.(\ref{sextic1}), under similar condition ($n=2$), the energy eigenvalues
can be derived from ,

 \begin{eqnarray}
 Q_{2}(\tilde{E})= {\tilde{E}}^2 -
 \frac{2\beta{\tilde{E}}}{\sqrt\gamma}-4n{\sqrt\gamma}=0 \qquad .
 \end{eqnarray}
The corresponding two solutions,
 \begin{eqnarray}
 E_{\pm} =\frac{3\beta}{2\sqrt\gamma}{\pm}{\left({\frac{{\beta}^2}{\gamma}+
 8\sqrt\gamma}\right)^{1/2}} \qquad ,
 \end{eqnarray}
 yield,
\begin{eqnarray}
\psi_{2_\pm}(x)=\exp{\left(-{\frac{{\beta}x^2}{4\sqrt\gamma}}-
{\frac{{\sqrt\gamma}x^4}{4}}\right)}P_{2_{\pm}}(x) \quad ;\\
{\mathrm here} \qquad
P_{2_{\pm}}(x)=1+{\left[\frac{\beta}{\sqrt\gamma}{\pm}{\left({\frac{{\beta}^2}{\gamma}+
 8\sqrt\gamma}\right)^{1/2}}\right]}x^2 \quad .
\end{eqnarray}

The case of $n=4$, with $\beta=0$ and for negative values of
$\alpha$, is a double-well potential. For
specificity, we take $\gamma=1$, which yields $\alpha=-11$ and
hence
\begin{eqnarray}\label{dbw} \hat{H} =
-\frac{d^2}{dx^2}-11x^2+x^6 \quad . \nonumber
\end{eqnarray}
 The energy eigenvalues for this Hamiltonian are obtained from,
$Q_{3}(E)= E(E^2-64)=0$, which gives $E= -8$, $0$ and $+8$. The
respective polynomial parts of the wave functions are,
\begin{eqnarray}
\label{dwel1}P_{4_{-}}(x)&=&1+4x^2+2x^4   \quad,\\
\label{dwel2}P_{4_{0}}(x)&=&1-(2/3)x^4   \quad,\\
{\mathrm and} \qquad \label{dwel3}P_{4_{+}}(x)&=&1-4x^2+2x^4
\quad .
\end{eqnarray}
It is clear that, the ground state has no nodes on the real line
and the subsequent two states have, respectively two and four
nodes on the real line, as desired. The above procedure
straightforwardly extends to higher values of $n$. We note that,
polynomial potentials, with the highest degree ${4m+2}$, with $m$
integer can also be solved in an analogous manner, provided
appropriate conditions are imposed on the couplings. It should
also be pointed out that, the solutions obtained so far,
correspond to the root $\lambda=0$. The other root $\lambda=1$,
does not lead to a closed form expression for the solution. Hence,
the odd states are not amenable to an analytical treatment.  This
is a manifestation of the QES nature of these quantal problems.

%%%%%%%%%%%%%%%%%%%%%%%%%%%%%%%%%%%%%%%%%%%%%%%%%%%%%%%%%%%%%%%%%%%%%%%%
\section{Approximation scheme for the analytically inaccessible states}

As is clear from the previous example, it is not possible to analytically
determine the states, other than the ones allowed by the condition given in
Eq.(\ref{condi}). In this section, we present an approximation scheme for
finding these {\em{analytically inaccessible states}}, taking the first excited
 state of the above mentioned double-well potential, as the example. Assuming
that, the measure is same for all the states, the residual part, which will be
suitably approximated to a finite degree polynomial in the following, should
have one zero on the real line. The rest of the zeros, if present, should lie
on the complex plane. In developing an approximation scheme, one also needs to
take care of the convergence of the series. The above two criteria are
utilized for finding out the approximate eigenvalue and eigenfunction of the
first excited state, which can then be extended to other states.

Expanding Eq. (\ref{sextic1}) for $\lambda=1$, one obtains,
 \begin{eqnarray} \label{exdbw} P_{4}^{1}(x) =
x-E\frac{x^3}{3!}+(E^2-36)\frac{x^5}{5!}
 +(76E-E^3)\frac{x^7}{7!}+ (E^4+8E^2-3024)\frac{x^9}{9!}+ \cdots
 \end{eqnarray}
This series does not terminate at any finite order. In developing an approximation
scheme, one possible way is to equate the coefficient of an appropriate term in
the series to
zero, in order to approximate it as a polynomial. This determines the energy
eigenvalue and eigenfunction; one also has to
show that higher order terms attain sufficiently small values to ensure convergence.

The other possible way is a variational approach. The fact that
our procedure of solving the differential equation yields a series
solution in terms of energy, allows us to treat the same as a
variational parameter. This can be effectively used for the
identification of the eigenstates and corresponding eigenvalues,
as will be shown below. Since the ground state is known
and we want to approximate the excited state as a product of the
ground state measure and a finite degree polynomial, we are left to
identify only the polynomial part of the excited state. This can
be carried out in a convergent manner to the desired degree of
accuracy. We first terminate the series given by Eq.(\ref{exdbw})
at some finite order. Now we check that this approximate
polynomial solution $u(x)$ has one zero on the the real line, this
gives us a range of energy values satisfying the above mentioned
condition. Since, we know that $u(x)$ is an approximate solution,
substitution of the same in the eigenvalue equation
$\tilde{H}u(x)=Eu(x)$ will leave a residual term. To get the best
approximate value for the energy at this order, we minimize
$\Delta$ defined as,

\begin{eqnarray}
\Delta =|\int_{-\infty}^{\infty} e^{-\frac{x^{4}}{2}}u^{*}(x)
(\tilde {H}-E)u(x)dx| \quad ,
\end{eqnarray}
using $E$, as a variational parameter.
For a better appreciation of the above points, we give below a plot of
$\Delta$ versus $E$ for a given approximate solution. Minimum of the plot indicates
the best approximate excited state for a given degree of the polynomial.
The approximate energy eigenvalue obtained by a higher degree polynomial gives
better result, as can be easily seen by comparison with the numerically
calculated value, given in Table I. This can also be noticed from Fig.1, which
clearly shows that, the minimum of $\Delta$ of the higher degree polynomial
occurs for a value of $E$ closer to the correct result. It should also be
noticed that the value of the residual term $\Delta$ is much smaller for the
higher order polynomial. Incorporation of higher order terms makes the
convergence better in the present scheme. It is found that, the first few
terms in our approximate solutions are enough to yield the excited state
eigenvalues, accurate upto the third place of the decimal, as shown in Table I.

\begin{center}
\bf{Table I. Comparison of approximate eigenvalues $E_{1}$ and $E_{2}$ for
the solutions of degrees five and nine respectively, with 
corresponding numerically calculated values $E_{Num}$ }
\end{center}
\begin{center}
\begin{tabular}{|c|c|c|c|c|c|}\hline

Energy
State&$E_{Num}$&$E_{1}$&Deviation&$E_{2}$& Deviation\\
\hline
 &&&&&\\
%$0^{*}$&$-8$&$--$&$--$&$ --$ &-- \\
%&&&&&\\
$1$&$-7.917350$&$-7.913704$&$0.04\%$ & $-7.916400$&$0.011\%$ \\
&&&&&\\
%$2^{*}$ & $0$ & $--$& $--$ &$--$&\\
%&&&&&\\
$3$ & $2.520359$
&$2.419229$&$4.01\%$ &$2.549348$&$1.15\%$\\
&&&&&\\
$5^{*}$ & $14.112964$ &$13.403590 $& $5.02\%$ &$--$& $--$\\
&&&&&\\
\hline
\end{tabular}
\end{center}
\begin{center}
$^{*}$ This eigenvalue corresponds to an approximate solution of degree
thirteen.
\end{center}
%------------------------Fig. 1-----------------------------

\begin{center}
\input epsf
\leavevmode{\epsfxsize=3.0in\epsfbox{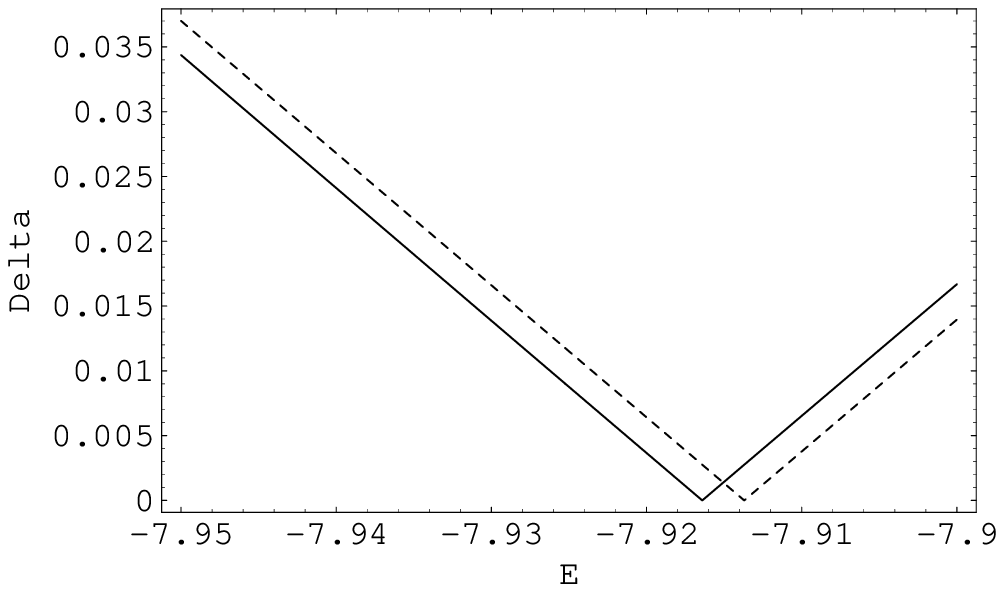}}
\end{center}
FIG. 1. Plots depicting the minima of $\Delta$ near the numerically evaluated 
eigenvalue, for the first excited state of the double-well potential.
 The solid and dashed curves show the variations of $\Delta$, when the
 approximate polynomial parts of the wave function are of degrees nine and
 five, respectively.

%----------------------------------------------------------- 
\vskip0.5cm
Apart from finding the states in between the known ones,
as has been done above, one can also
approximate the even and odd states lying above the analytically determinable
part of the spectrum. For this purpose, we plot $\Delta$ for odd and
even states in a range of energy values, in Figs. 2 and 3 respectively. This is
computed taking approximate solution of degree nine for the odd states;
and twelve for the even ones.
%------Fig2 &3
%\newpage
\begin{center}
\input epsf
\leavevmode{\epsfxsize=3.0in\epsfbox{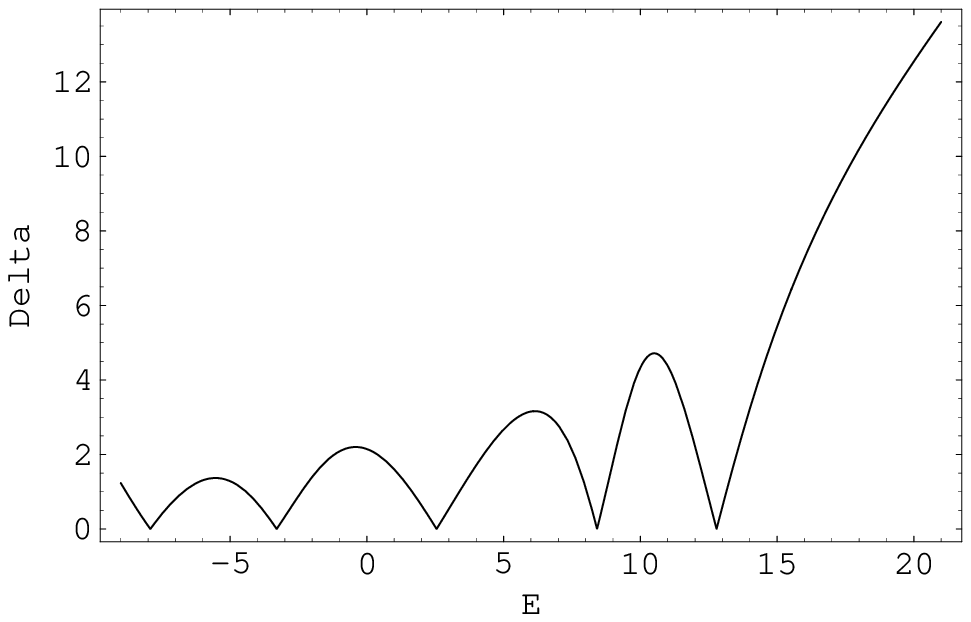}}
\end{center}
FIG. 2. Variations of $\Delta$ with respect to $E$, in the odd-parity
sector. Here the polynomial part of the wave function has degree nine and
the energy range is much bigger, as compared to Fig.1. One clearly notices
that several minimum values of $\Delta$ occur closer to the numerically
determined eigenvalues.
\begin{center}
\input epsf
\leavevmode{\epsfxsize=3.0in\epsfbox{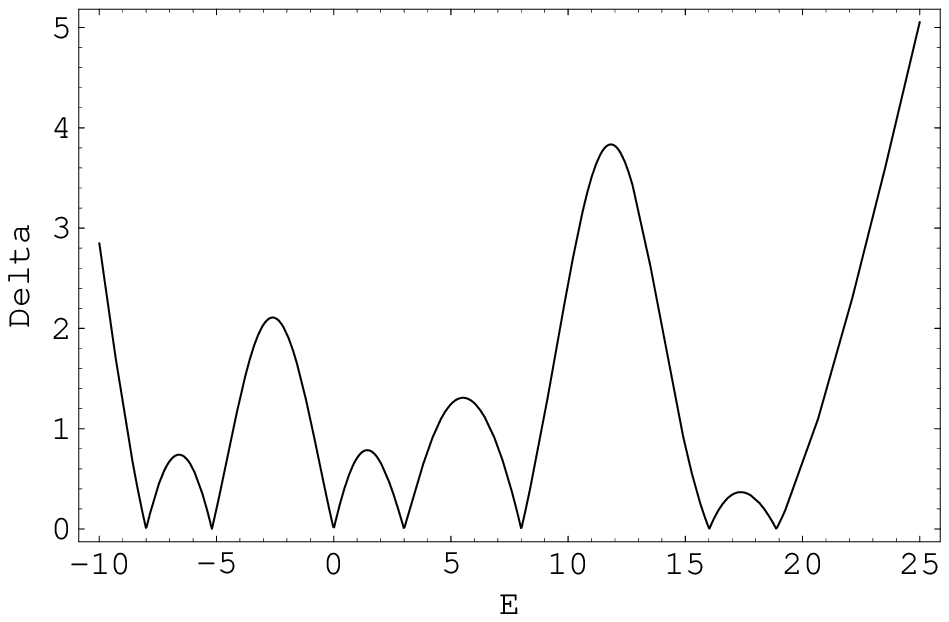}}
\end{center}
FIG. 3. Plot of $\Delta$ versus $E$, for the even states, when the approximate
polynomial solution is of degree twelve. Like the odd sector in Fig.2, here
also some of the minima are located close to the corresponding numerical
values.
\vskip0.5cm

%-----------------
It is extremely interesting to observe that, various minima of the
plot show the expected locations of the states in the energy space, with
reasonable accuracy, as seen from Table II., containing the numerically
calculated eigenvalues of higher excited states. It is
worth pointing out that, all the minima of the plot do not correspond to
physically acceptable states. One needs to take those values, which fulfill the
required conditions on the wave functions. At this stage of approximation, the
computed eigenvalues for the higher excited states agree with the numerically
obtained ones upto $5\%$ accuracy. As has been done earlier, in order to
improve the result further, one
needs to take still higher degree polynomials.
%\newpage

Below, we give another table containing numerically calculated energy
eigenvalues of higher excited states of the double-well potential.
It is clear from the spectrum, that higher excited states are approximately equispaced as
expected.
%\newpage
%*************************Table 2 *********
\begin{center}
{\bf{Table II. Numerically calculated energy eigenvalues of a number of
excited states of the double-well potential}}
\end{center}

\begin{center}
\begin{tabular}{|c|c|c|c|}\hline

Energy State&$E_{Num}$&Energy State&$E_{Num}$\\
\hline &&&\\
$6$&$21.1575028$&$13$& $89.0540515$ \\
&&&\\
$7$&$28.9747742$&$14$&$100.9447956$ \\
&&&\\
$8$ & $37.4938424$ & $15$ & $113.3093212$ \\
&&&\\
$9$ & $46.6606249$ & $16$& $126.1303320$ \\
&&&\\
$10$ &$56.4324169$ & $17$ & $139.3922937$ \\
&&&\\
$11$ &$66.7742029$ & $18$ & $153.0811528$ \\
&&&\\
$12$ &$77.6565110$ & $19$ & $167.1841151$ \\
&&&\\
\hline
\end{tabular}

\end{center}

%\newpage
%**********************************************************

\section{Conclusions}
 In conclusion, we have applied a recently developed scheme for solving linear
 differential equations, to find the exact eigenstates of a wide class of
 quasi-exactly solvable Hamiltonians and also to develop a convergent approximation
 scheme to determine the states not obtainable by analytical treatments. The
 fact that the method used provided an expansion of the wave functions in terms
 of the corresponding energies, allowed us to treat energy as a variational
 parameter, in the approximation scheme. An independent numerical scheme
 was used to check that the low-lying eigenvalues are accurate up to third place of
 decimal, with only a few terms from the series expansion. The higher excited
 states need more number of terms, since these have $5\%$ accuracy, under
 similar conditions. It also needs to be pointed out that, the QES systems treated here are equipped
with a ground state measure, which facilitates the working of the approximation
scheme. It will be interesting to extend the present scheme to other
 anharmonic potentials.

 {\bf{Acknowledgement :}} We acknowledge many useful discussions
with Prof. V. B. Sheorey, who has also made available to us the
algorithm for the numerical solution of the QES eigenvalue
problems.

\end{document}